# Magnetic properties of cobalt doped $ZrO_2$ nanoparticles: Evidence of Co segregation

Francisco González Pinto[1, Δ], Mariano Andrés Paulin[2,3, Δ,*], Ana Gabriela Leyva[1,3,4], Joaquín Sacanell[1,3]

[1]Depto. de Física de la Materia Condensada, Gerencia de Investigación y Aplicaciones, Centro Atómico Constituyentes, CNEA, Av. Gral. Paz 1499, San Martín (1650), Buenos Aires, Argentina.

[2]Div. de Resonancias Magnéticas, Gerencia de Física, Centro Atómico Bariloche, CNEA, Av. Bustillo 9500, San Carlos de Bariloche (8400), Rio Negro, Argentina.

[3]INN, CNEA-CONICET, Av. Gral. Paz 1499, San Martín (1650), Buenos Aires, Argentina.

[4]Escuela de Ciencia y Tecnología, Martin de Irigoyen 3100, Edificio Tornavía, Campus Miguelete, UNSAM, San Martín (1650), Buenos Aires, Argentina.

*Corresponding author: Dr. Mariano Andrés Paulin
Depto. de Resonancias Magnéticas, Gerencia de Física, Centro Atómico Bariloche, Av. Bustillo 9500, San Carlos de Bariloche (8400), Rio Negro, Argentina,
+5402944445120, mariano.paulin@cab.cnea.gov.ar

[Δ] Both authors equally contributed to this work.

## Abstract

We synthesized pure and Co-doped (6.25 - 12.5% at.) $ZrO_2$ nanopowders in order to study their magnetic properties. We analyzed magnetic behavior as a function of the amount of Co and the oxygenation of the samples, which was controlled by low pressure thermal treatments. As prepared pure and Co-doped samples are diamagnetic and paramagnetic respectively. Ferromagnetism can be induced by performing low pressure thermal treatments to the samples, which becomes stronger as the dwell time of the thermal treatment is increased. This behavior can be reversed, recovering the initial diamagnetic or paramagnetic behavior, by performing re oxidizing thermal treatments. Also, a cumulative increase can be observed in the saturation of the Magnetization with the number of low pressure thermal treatments performed to the sample. We believe that this phenomenon indicates that cobalt segregation induced by the thermal treatments is the responsible for the magnetic properties of the $ZrO_2$-Co system.

**Introduction**

The development of spintronics has promoted the study of materials in which magnetism and electrical transport are correlated. Since the theoretical proposal of Dietl et al. of room temperature ferromagnetism in Mn doped ZnO [1], a significant amount of work has been devoted to the study of the magnetic properties of doped and undoped oxides such as ZnO, $TiO_2$, $SnO_2$, $HfO_2$ and $CeO_2$ [2,3,4,5]. Among them is $ZrO_2$, for which room temperature ferromagnetism has been predicted in Mn doped $ZrO_2$ by *ab initio* calculations [6]. Since that, several works appeared exploring $ZrO_2$ as a candidate for a diluted magnetic oxide (DMO).

In its pure state, the $ZrO_2$ system is generally diamagnetic or paramagnetic, but when doped with magnetic elements (mostly Mn and Fe) the compound can behave as a paramagnet or a ferromagnet [7,8,9,10,11,12]. Also, the absence of magnetic behavior was reported [13,14]. Even more, the appearance of ferromagnetism in $ZrO_2$, was shown to be superimposed with a resistive switching induced by the application of electric pulses [15]. That broad spectrum of results shows that, even the topic of DMO has been studied in detail, a complete understanding of their magnetic properties, in particular for the $ZrO_2$ system are far from being clear. Also, despite numerous claims of successful growth of DMOs, few works have convincingly demonstrated the synthesis of real, intrinsic DMO and the question of their real utility remains open.

The presence of oxygen vacancies has been ascribed to play a significant influence on the magnetic properties of several DMO [4,10,13]. For that reason, much recently, research of $ZrO_2$-based DMOs has focused on the study of nanostructures [16,17,18,19,20,21] with large surface/volume ratio and thus a proper environment for oxygen vacancies.

In several of those studies, oxide segregation is mentioned as responsible for the magnetic behavior but an analysis of this problem is only performed in few of them [21,22]. This is due to the typically small magnetic signals of DMOs, which can arise by an amount of segregated magnetic element below the detection limits of the typical laboratory techniques [23], making very difficult to identify segregation through the use of X ray diffraction or adsorption techniques.
Magnetization, on the other hand, is extremely sensitive to small amounts of magnetic impurities, being its drawback that we are dealing with a bulk technique. However, as we will show, the use of systematic experiments varying oxygenation of the samples, can be used to discard the presence of impurities and also to evidence the influence of segregation.

In this work we study the magnetic properties of $ZrO_2$ nanoparticles doped with Cobalt as a function of doping and the relative concentration of oxygen vacancies. We have chosen the Co-$ZrO_2$ system because is one of the less explored material as DMO. We use nanoparticles instead of thin films for two main reasons: First, the typically low magnetic signals of these systems can be easily enhanced just by increasing the amount of sample. Second, as mentioned above, the surface to volume ratio of the sample can be adjusted simply by performing adequate thermal treatments, which is particularly convenient as the surface is usually the preferred location of point defects such as oxygen vacancies.

We will contribute to answer on whether magnetism is an intrinsic property of the oxide or if it is due to segregation of the dopant, as there is still controversy regarding this aspect. To discriminate the origin of the observed magnetic signal, we performed a series of systematic experiments. The experiments are combinations of series of low pressure thermal treatments and thermal treatments in air. This was done, as in other systems [5,24,25], in order to study the relative influence of oxygenation on the magnetic properties of the samples, and also to test reversibility against oxygen incorporation. The choice to rely mainly on magnetization data instead of other techniques, is due to the fact that the amount of segregated material that could explain the small magnetic signals observed in DMOs is typically below the detection limit of the most common laboratory techniques. The systematic experiments allowed us to discriminate among intrinsic and extrinsic (impurities, segregation) contributions.

Our results indicate that magnetism is increased in de-oxygenated samples and it can be reversed by performing thermal treatments in an oxidizing atmosphere. On the other hand, the application of cumulative thermal treatments at low pressure and air, results in an increase of the ferromagnetic signal, thus indicating that segregation is the responsible for the ferromagnetic behavior.

**Experimental**

Pure and Co-doped $ZrO_2$ (6.25%, 9% and 12.5% at.) powders were synthesized by the liquid-mix method using $ZrO(NO_3)_2 \cdot 2H_2O$ and Cobalt(II) nitrate hexahydrate as reagents. All samples were dried at 120°C for 72 hours, grounded and annealed in air at 500°C for 2 hours.

Given that the presence of defects, in particular oxygen vacancies, is considered to play a key role in the magnetic behavior of DMOs, we performed a series of low pressure thermal treatments (LPTT), at approx. $3 \cdot 10^{-2}$ mbar. The degree of de-oxygenation was qualitatively controlled by varying the dwell time (dwt) of the LPTT. We also performed thermal treatments in air (TTA) in order to oxidize the samples. Cycles consisting of sequences of LPTT and TTA treatments were used in order to study cumulative effects. Both treatments were performed at 450°C, which is lower than the synthesis temperature, in order to avoid changes in the crystallite size of the samples.

X-ray powder diffraction patterns of the samples were obtained using a PANalytical EMPYREAN diffractometer. The pure samples showed the formation of tetragonal $ZrO_2$ with a small fraction of monoclinic $ZrO_2$. In Co-doped samples, however, just the tetragonal structure was observed with no appreciable secondary phases associated with the dopant. Crystallite sizes of around 7 nm were obtained for all samples by means of the Scherrer equation. No changes in the composition of the samples or the crystallite size were observed upon performing the LPTT's, as verified by EDS and X-ray diffraction respectively.

Magnetic measurements of powdered samples, were performed in a Versalab[TM] Vibrating Sample Magnetometer from Quantum Design at 300 K. The signal corresponding to the sample holder was always measured and subtracted from the results.

**Results and Discussion**

In figure 1 we show the XRD data for the undoped samples. Results for samples thermally treated at low pressure for 0, 4, 8 and 12 h. are presented. The presence of a mainly tetragonal phase has been identified, with a minor contribution of a monoclinic phase. The stabilization of these phases is usually observed in nanostructured samples[26,27,28]. No significant difference is observed for the samples subjected to low pressure thermal treatments.

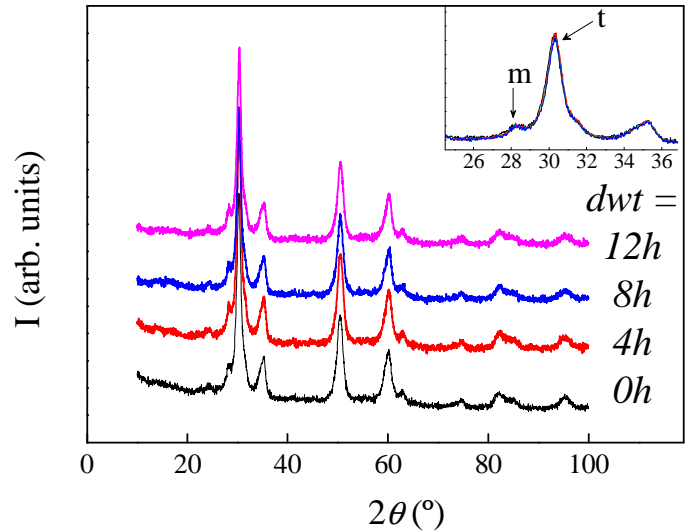

**Figure 1:** X ray diffraction data for $ZrO_2$ undoped samples subjected to low pressure thermal treatments from 0 to 12 h as described in the experimental section. Inset: detail of the most significant reflections of the tetragonal (t) and monoclinic (m) contributions for alls samples.

The XRD data corresponding to samples doped with Co ($x \approx 0 - 12.5\%$) is presented in figure 2(a). From these data, the presence of $ZrO_2$ is clear, and no evidence of segregation of dopant nor secondary phases can be observed. This indicates that most part of the Co ions have entered into the crystalline structure of the oxide and segregation, if any, is below the detection limit of our XRD data. We performed similar low pressure thermal treatments to the doped samples, as those presented in figure 1 for the pure $ZrO_2$ compound. We can see in figure 2(b) that no structural changes are induced in this case also (shown for the sample doped with 12.5% of Co).

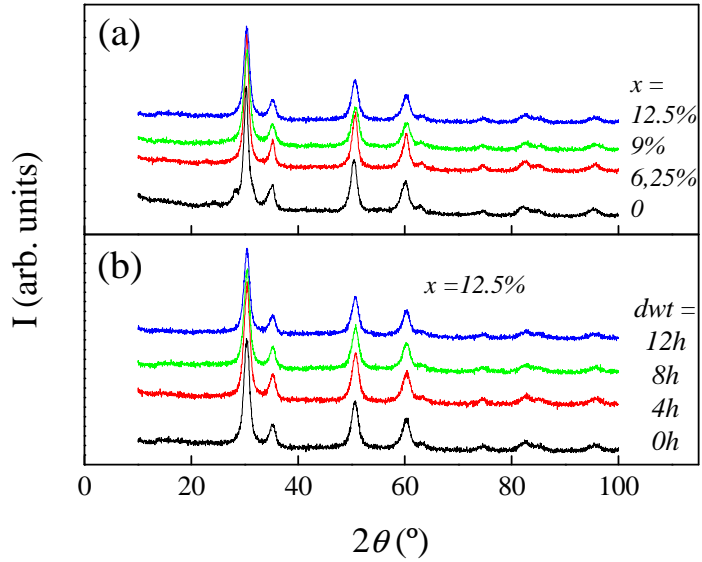

**Figure 2:** (a) X ray diffraction data (a) for $ZrO_2$ doped samples, (b) for samples with 12.5% of Co, subjected to low pressure thermal treatments from 0 to 12 h as described in the experimental section.

In table I we present the crystallite size for several samples as obtained using the Scherrer equation. We can see that no significant changes can be observed both on doping and with the low pressure thermal treatments, even though a slight reduction of the crystallite diameter is observed on samples with doping above 9% of Co.

| τ (h) | 6.25% of Co | 9% of Co | 12.5% of Co |
|---|---|---|---|
| 0 | 8 ± 1 nm | 6.2 ± 0.5 nm | 6.2 ± 0.7 nm |
| 4 | 8 ± 1 nm | 6.2 ± 0.5 nm | 6.2 ± 0.8 nm |
| 12 | 8 ± 1 nm | 6.5 ± 0.7 nm | 6.5 ± 0.7 nm |

**Table I:** crystallite diameter for several samples as a function of %Co and the time of the low temperature thermal treatment in hours (h).

In figure 3 we show the magnetization ($M$) as a function of the magnetic field at room temperature, for samples synthesized in air, with x = 0, 6.25, 9 and 12.5%. We see that the behavior ranges from diamagnetic for the non doped sample, to paramagnetic for the doped ones, with a progressively increased paramagnetic susceptibility while increasing the concentration of the dopant.

In the inset of figure 3 we show the values of $M$ for 2T, extracted from the $M$ vs $H$ data and displayed as a function of doping. We see an almost linear increase of the magnetization with a slight tendency towards saturation suggesting a solubility limit reached at around 12,5% of Co concentration.

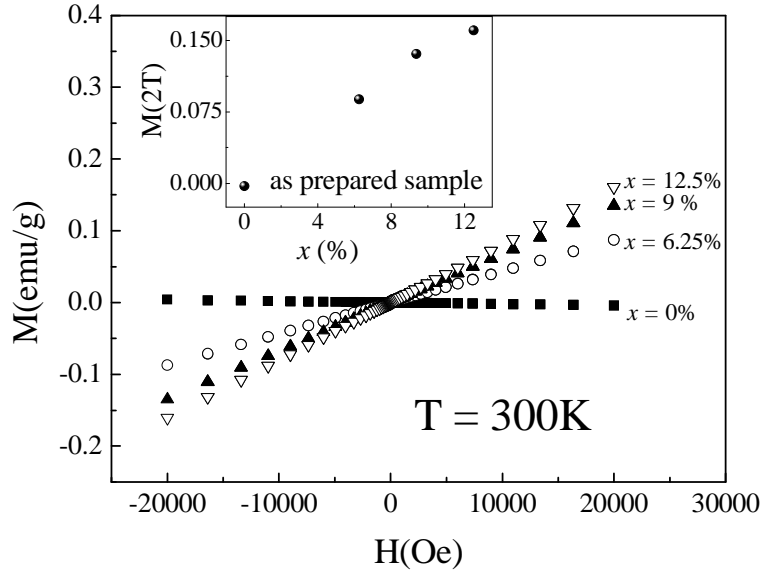

**Figure 3:** Magnetization as a function of magnetic field for the as-prepared samples. Doping is indicated in the figure. Inset: Magnetization evaluated at H = 2 Tesla as a function of doping.

To evidence the role played by oxygen on the magnetic properties of the system, we have performed a series of thermal treatments in vacuum to de-oxygenate the samples. The color of the samples changed upon performing the LPTT's. We observed a darkening of the samples that strengthened when increasing the dwell time of the treatment, indicating changes in the optical band-gap and therefore in the electronic structure of the system.

Magnetic susceptibility data was analyzed with the Curie Law, corresponding to a paramagnetic material (eq. 1) [29].

$$\chi = \frac{N}{V} \frac{(\mu_B p)^2}{3 k_B T} \qquad (1)$$

Where $\mu_B$ is the Bohr magneton and $p$ is the effective number of Bohr magnetons. Assuming that N/V is the number of Co atoms per unit volume, we obtained $p$ by fitting the magnetic susceptibility data and is presented in table II. The increment in the values of $p$ are consistent with an increase in the oxidation state of Co, taking into account that the spin only contribution (high spin) of $Co^{2+}$, $Co^{3+}$ and $Co^{4+}$ are 3.87, 4.9 and 5.92, respectively.

|                  | 6.25%         | 9%          | 12.5%          |
|------------------|---------------|-------------|----------------|
| $p$ (as prepared)| 4.62 ± 0.02   | 4.6 ± 0.1   | 4.47 ± 0.05    |
| $p$ (LPTT 4h)    | 4.8 ± 0.1     | 4.8 ± 0.1   | 4.8 ± 0.2      |
| $p$ (LPTT 12h)   | 5.0 ± 0.2     | 5.7 ± 0.5   | 6 ± 1          |

**Table II:** Effective number of Bohr magnetons for samples doped with 6.25, 9 and 12.5% of Co, extracted from magnetic susceptibility data.

The magnetization as a function of magnetic field for de-oxygenated samples at 300 K is shown in figure 4. When subjected to a LPTT of 4 hours, magnetization measurements show little change for undoped and low doped samples with respect to the as-prepared ones but hysteresis loops have been observed for samples with 9 % and 12.5% of Co doping. We can see a progressive enhancement of the ferromagnetic behavior while increasing Co concentration. However, it is worth to note that a paramagnetic contribution is present in all samples (for x > 6.25%, it is superimposed to the FM contribution). This component varies on doping in a similar way as that presented by the as-prepared samples. The values of *p* obtained for the samples subjected to LPTT display an increase that suggests a slight tendency of Co towards the $Co^{2+}$ oxidation state (see table II). The formation of $Zr^{3+}$, would also lead to an increase of the value of p, however this is far less likely than the partial reduction of Co ions.

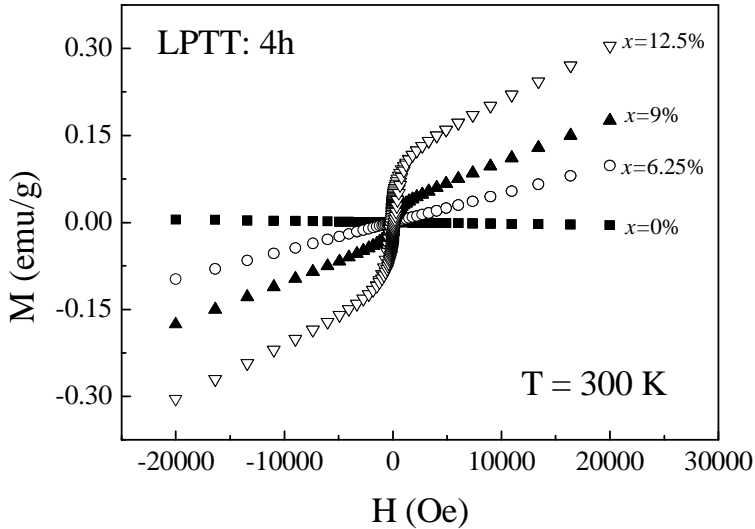

**Figure 4:** FM hysteresis loops observed for the samples after performing a low pressure thermal treatment of 4 h.

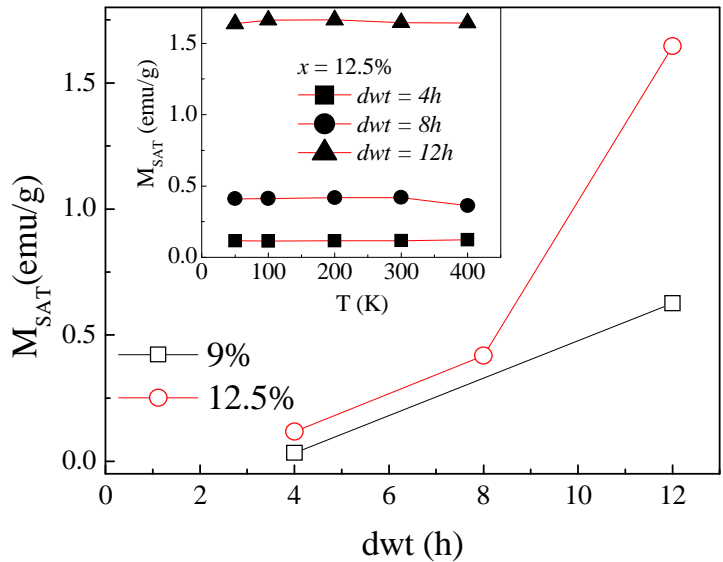

**Figure 5:** Saturation value of the magnetization ($M_{SAT}$) as a function of the dwell time of the low pressure thermal treatment for samples with 9% and 12.5% doping. Inset: $M_{SAT}$ as a function of temperature for the sample with x=12.5% subjected to low pressure thermal treatments o 4, 8 and 12h.

The saturation value of the FM component of the magnetization ($M_{SAT}$) increases both with doping (see fig. 4) and with the dwell time of the LPTT, as shown in figure 5. It also can be seen that is independent of temperature. The observed increment of $M_{SAT}$ vs. dwt is clearly correlated with de-oxygenation, thus qualitatively supporting the idea that an increase in the concentration of oxygen vacancies is essential to explain the appearance of ferromagnetism. However, it is insufficient to discard other sources of FM, such as clustering of Co (both as metallic or as a magnetic oxide).

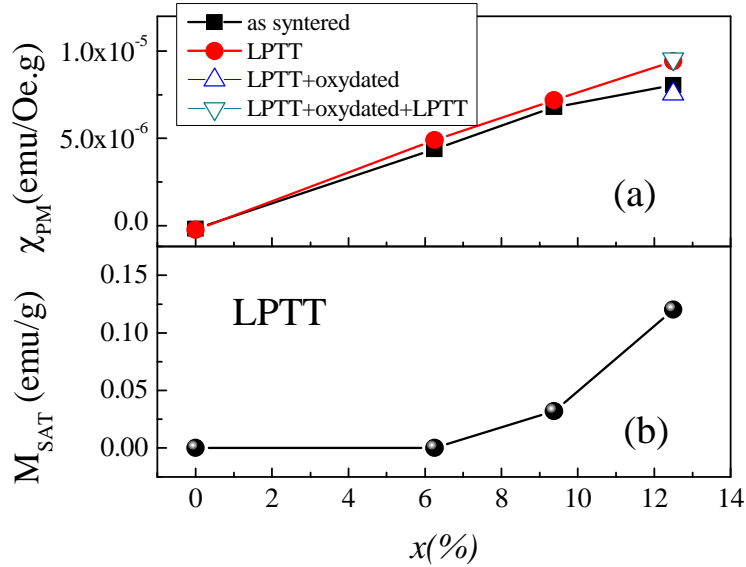

**Figure 6:** (a) Paramagnetic susceptibility as a function of doping after performing several thermal treatments in different atmospheres. (b) $M_{SAT}$ for a sample after performing a low pressure thermal treatment of 4 h at 450ºC. Measurements performed at room temperature.

In figure 6(a) we show that the paramagnetic susceptibility ($\chi_{PM}$) for large magnetic field (extracted from the data of figures 3 and 4), displays an increment with doping. We can see that the paramagnetic contribution of the as-prepared sample and the de-oxygenated sample (LPTT of 4 h) show a similar trend as a function of doping until 9%. Thus, this component is not significantly affected by oxygenation, at least for low doping. However, for $x = 12.5\%$, we can see several interesting features: 1) The as-prepared sample displays lower $\chi_{PM}$ that the sample subjected to an LPTT of 4h, 2) $\chi_{PM}$ of the sample that was previously subjected to an LPTT is reduced if the sample is then oxygenated (up triangle, indicated as LPTT+oxy), reaching almost the same value as the as-prepared sample and 3) if a subsequent LPTT is applied to this sample, $\chi_{PM}$ recovers its initial value (down triangle, indicated as LPTT+oxy+LPTT in figure 6(a)).

In figure 6(b) we show the value of the saturation of the magnetization ($M_{SAT}$) of the samples subjected to a LPTT of 4h at 450ºC as a function of doping (the PM contribution was previously subtracted). A monotonous increase can be observed on doping. This dependence, along with the aforementioned change in color of the de-oxygenated samples, evidences a correlation between a change in the electronic structure and the appearance of ferromagnetism. Both changes undoubtedly arise due to the some kind of de-oxygenation of the samples.

Thus, in the case of low doped samples, the LPTT mainly affects ferromagnetism, while for high doping, where segregation is more likely to occur, the paramagnetic contribution is also affected. This non trivial behavior could be related with an enhanced segregation in the highly doped samples. In that case, the presence of different cobalt oxides with lower $p$ values is possible, as cobalt ions can be in intermediate situations between non-interacting (PM), in a weak AFM interaction or strongly interacting (FM).

Inspired by the reversible behavior of the susceptibility of the sample doped with 12.5% of Co, shown in figure 6(a), we performed a sequence of M vs. H loops to that sample in the following sequence: 1) as prepared, 2) as prepared + subjected to an LPTT of 4h and 3) as prepared + subjected to an LPTT of 4h + further oxygenated. In figure 7 we show the M vs H measurements obtained for that sequence. High field slopes are similar for all measurements, the most significant difference being the appearance of ferromagnetism after the LPTT is performed. However (as anticipated in figure 6(a)) the paramagnetic component, which is superimposed to the ferromagnetic hysteresis cycle, presents a slight increase from the as-prepared sample to the one that is subjected to an LPTT.

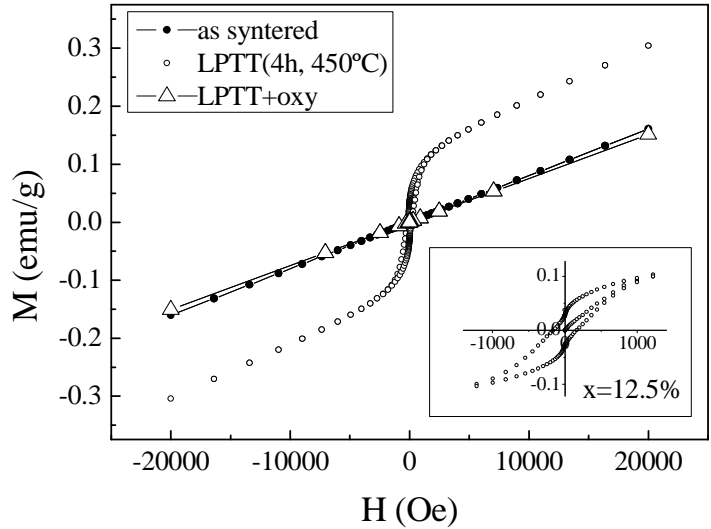

**Figure 7:** M vs. H dependence for the as prepared sample with $x$ =12.5% doping. A reversible behavior can be observed on performing low pressure and oxygenating thermal treatments in air. Inset: Detail of the hysteresis loop at low magnetic field. Measurements performed at room temperature.

The most significant effect is a switching behavior against reduction/oxidation of the samples, in which an initially PM sample turns into FM and then goes back to PM. This finding is similar to what can be observed in systems like Co doped ZnO [25] and $CeO_2$ [5], showing that FM behavior is not the result of the accidental incorporation of spurious magnetic impurities during thermal treatments. We observed that both the appearance of the hysteresis loop and the change in color are reverted when re-oxidizing the sample. We also see that the increase of the paramagnetic susceptibility mentioned in the previous paragraph is reversed.

Within this scenario, two possibilities arise to explain our results:
- FM is an intrinsic property of a homogeneous compound which depends on the amount of oxygen vacancies generated during the LPTT.

- FM is the result of segregation of magnetic material induced by the LPTT.

The first possibility is clear and in the second, the obtained values of saturation magnetization could correspond to the presence of an amount of a magnetic compound of cobalt, so small that it cannot be detected in an XRD analysis, as has been shown in other systems [30,31].

To validate the "segregation" possibility, we performed a number of successive reducing/oxidizing LPTT – TTA cycles on the 12.5% doped sample. It is clear that, if ferromagnetism is an intrinsic feature or it arises from (static) segregation induced during the synthesis procedure, the situation will resemble the data of figure 5, in the sense that after each LPTT the magnetization should go "back and forth" between the PM and the FM state. If, on the other hand, segregation is assisted by the thermal treatments, a cumulative behavior should be observed after each LPTT-TTA cycle is performed.

In figure 8(a), we present M vs. H curves measured after performing 1, 2, 3 and 4 LPTT to the sample. We can see that the saturation of the FM part of the magnetization measured increases as a function of the number of LPTTs.
Each LPTT was followed by a TTA, after which the paramagnetic behavior was recovered as shown in figure 8(b). In figure 8(c) we see that after each cycle, $M_{SAT}$ increases exponentially, growing almost two orders of magnitude after 4 thermal treatments.

To clarify our image: before the LPTT the sample is in its "as prepared" state, ferromagnetism is not observed because the segregated phase (if present) can be as a non-magnetic Co oxide ($Co_3O_4$ for example). This oxide tends towards metallic Co upon reduction and consequently, the FM behavior appears after each LPTT. Our results also show that after each LPTT-TTA cycle, the system evolves, as clearly seen in figure 8(a). This cumulative result suggests that the observed ferromagnetic behavior is a consequence of a segregated phase, promoted in principle by both (LPTT and TTA) thermal treatments.
Segregation can be induced in the whole LPTT-TTA cycle but, as (close to) metalic Cobalt is needed for FM behavior, ferromagnetism is only to be expected when measuring the sample after an LPTT is performed, as shown by our results.

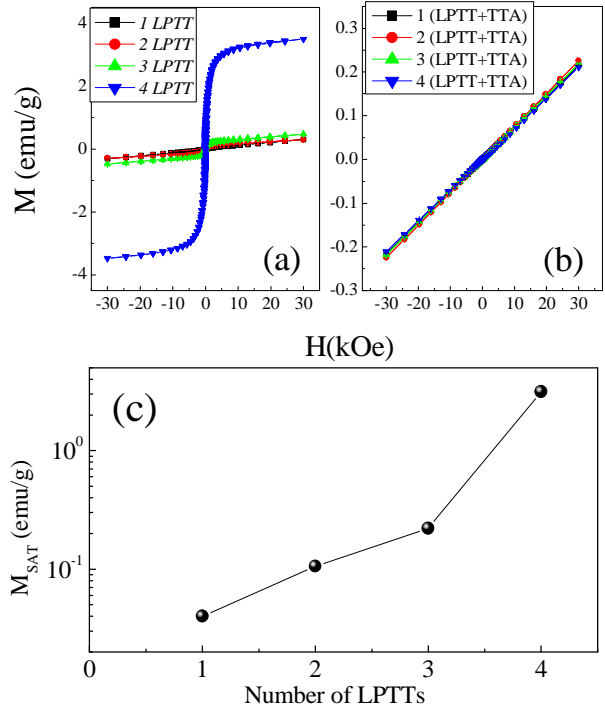

**Figure 8:** (a) Magnetization loops performed after 4 subsequent LPTT were performed to the sample with $x = 12.5\%$ doping. (b) Magnetization loops performed after 4 subsequent LPTT+TTA were performed to the sample with $x = 12.5\%$ doping. (c) Value of the saturation of the magnetization as a function of the number of LPTT performed to the sample.

**Conclusions**

In summary, we have shown that magnetic behavior can be induced by introducing Co in the $ZrO_2$ oxide. In the as prepared samples, only paramagnetism can be observed, while ferromagnetism appears if the samples are subjected to low pressure thermal treatments.

We performed experiments in which we alternated low pressure and ambient pressure thermal treatments, that showed that the origin of the ferromagnetic behavior is more likely to be related with cobalt segregation induced during those thermal treatments, thus indicating that ferromagnetism in the $ZrO_2$-Co system is due to Co segregation. We showed that DC magnetization, although it does not give a microscopic picture, it is capable to identify very small contributions that are undetectable by typical laboratory techniques, and its results can be useful if systematic studies are performed.

Even though ferromagnetism is not an intrinsic property of the oxide, the study of the magnetic properties of doped $ZrO_2$ still deserves attention as an heterogeneous insulator with magnetic aggregates.

An interesting theoretical proposal that, to the best of our knowledge, hasn't been experimentally tested yet is to induce magnetism in $ZrO_2$ by doping with non magnetic impurities such as K and Na [32]. Also, an experimental study on K doped $TiO_2$ [33] show, however, that weak paramagnetism is observed but no long-range ferromagnetic order. Another interesting aspect, is the analysis of the magnetic behavior in N-doped $ZrO_2$, as in very recent theoretical studies, ferromagnetism is suggested to appear in certain conditions [34], while its absence at room temperature is also claimed [35]. Besides, as small Co or Co oxide precipitates form part of this system, the dynamic behavior can be also a subject that deserves to be studied. In fact, superparamagnetic or spin glass-like behavior was observed in Fe doped $ZrO_2$ [19], which can be a consequence of the presence of magnetic clusters within a non-magnetic matrix. Therefore, further research on ferromagnetism in $ZrO_2$ based systems still deserves attention both from the basic and the technological point of view.

**Acknlowledgements**

Financial support from CONICET (PIP00038 and PIP00362) and ANPCyT (PICT 1327 and 1506) is acknowledged.